\documentclass[prd,showpacs,floatfix,nofootinbib,preprintnumbers]{revtex4}
\usepackage{epsfig}

\let\jnfont=\rm
\def\np#1,{{\jnfont Nucl.\ Phys. }{\bf #1},}
\def\pl#1,{{\jnfont Phys.\ Lett. }{\bf #1},}
\def\prd#1,{{\jnfont Phys.\ Rev.\ D }{\bf #1},}
\def\prl#1,{{\jnfont Phys.\ Rev.\ Lett.\ }{\bf #1},}
\def\sjnp#1,{{\jnfont Sov.\ J.\ Nucl.\ Phys. }{\bf #1},}
\def\zp#1,{{\jnfont Zeit.\ Phys.}{\bf #1},}

\def\bxll{B \rightarrow X_s \, \l^+ \, \l^-}

\def\beq{\begin{eqnarray}}
\def\eeq{\end{eqnarray}}
\def\nnb{\nonumber}
\def\aB0{\big| B^0 \big>}
\def\bB0{\big| \bar{B}^0 \big>}

\newcommand{\be}{\begin{equation}}
\newcommand{\ee}{\end{equation}}
\newcommand{\bea}{\begin{eqnarray}}
\newcommand{\eea}{\end{eqnarray}}
\newcommand{\ba}{\begin{array}}
\newcommand{\ea}{\end{array}}

\begin{document}
\title{CP violation in $B_{d,s} \to l^+l^-$ in the model III 2HDM}
\author{\bf{Yuan-Ben Dai,  Chao-Shang Huang, Jian-Tao Li, Wen-Jun Li}}

\address{ Institute of Theoretical Physics, Academia Sinica, Beijing 100080, China}

\date{\today}

\vspace{0.5cm}

\begin{abstract}
We have calculated the Wilson coefficients $C_{10}, C_{Q_i}$
(i=1,2) in the $\overline{MS}$ renormalization scheme in the model
III 2HDM. Using the obtained Wilson coefficients, we have analyzed
the CP violation in decays $B^0_q\rightarrow l^+l^-$ (q=d,s) in
the model.  The CP asymmetry, $A_{CP}$, depends on the parameters
of models and $A_{CP}$ in $B_d\rightarrow l^+l^-$ can be as large
as $40\%$ and $35\%$ for $l=\tau$ and $l=\mu$ respectively. It can
reach $4\%$ for $B^0_s$ decays. Because in SM CP violation is
smaller than or equal to O($10^{-3}$) which is unobservably small,
an observation of CP asymmetry in the decays $B^0_q \to l^+l^-
(q=d,s)$ would unambiguously signal the existence of new physics.
\end{abstract}

\pacs{11.30.E, 13.20.H, 12.60.F, 12.60.J}

\maketitle
\indent
\section{\bf Introduction}
The flavor changing neutral current process, $B_{d,s} \to l^+ l^-$
(l=$\mu,\tau$), has attracted a lot of attention since it is very
sensitive to the structure of SM and potential new physics beyond
SM and was shown to be powerful to shed light on the existence of
new physics before possible new particles are produced at
colliders~\cite{hly,cg,bk,csb,abbt}.
For example, in a very large region of parameter space
supersymmetric (SUSY) contributions were shown to be easy to
overwhelm the SM contribution\cite{hly,cg,bk,csb,abbt} and even
reach, e.g., for l=$\mu$, the experimental upper bound\cite{data}
\beq
B_r (B_d \rightarrow \mu^{+}\mu^-) < 6.8 \times 10^{-7} \ \ (CL=90\% ) \nnb \\
B_r (B_s \rightarrow \mu^{+}\mu^-) < 2.0 \times 10^{-6} \ \
(CL=90\% ). \label{limit} \eeq In other words measuring the
branching ratio of $B_{d,s} \to l^+ l^-$ can give stringent
constraints on the parameter space of new models beyond SM,
especially for that of the minimal supersymmetric standard model
(MSSM) because of the tan$^3\beta$ dependence of SUSY
contributions in some large tan$\beta$ regions of the parameter
space\cite{hly,hpt,bk,csb}. In the model II two Higgs doublet
model (2HDM), the branching ratio of $B_s \rightarrow
\mu^{+}\mu^-$ can also reach $2\times 10^{-8}$, which is within
good reach at Tevatron Run II, if $\tan\beta$ is large enough
(say, $\geq 60$ )~\cite{ln,csb}. Comparing with hadronic decays of
B mesons, this process is very clean and the only nonperturbative
quantity involved is the decay constant that can be calculated by
using the lattice gauge theory, QCD sum rules etc.

An observable without hadronic uncertainty at all is the CP
asymmetry in  the process $B_{d,s} \to l^+ l^-$ since the common
uncertain decay constant, which is the only source of hadronic
uncertainty for the process, cancels out. CP violation in the
$b$-system has been established from measurements of
time-dependent asymmetries in $B \to J/\Psi K$ decays
\cite{sin2betababar,sin2betabelle}. It is urgent and important to
study CP violation in more processes, including the process
$B_{d,s} \to l^+ l^-$ which might be measured in the near future.
Obviously for the process $B_{d,s} \to l^+ l^-$  there are no
direct CP violations since there are no strong phases in the decay
amplitude. But it is well known that CP violating effects can
survive after taking into account the mixing of the neutral
mesons, $B^0$ and $\bar{B}^0$, in the absence of the strong
phases. Recently, it has been shown that CP violation in $B_{d,s}
\to l^+ l^-$ (l=$\mu, \tau$) is also an interesting observable for
searching for new physics~\cite{hl1,hl2}. In the present paper we
study it in the model III 2HDM~\cite{cs,m3}. It is well-known that
in the model III 2HDM the couplings involving Higgs bosons and
fermions can have complex phases, which can induce CP violation
effects, even in the simplest case in which all tree-level FCNC
couplings are negligible.  So it is expected that a significant CP
asymmetry in $B_{d,s} \to l^+ l^-$ (l=$\mu, \tau$) should exist
due to the effects of such extra phases in the model. Another
motivation to consider the model III is that one starts in the
minimal supersymmetric standard model (MSSM) with a Higgs sector
of type II 2HDM and, after integration of squarks and gluinos, one
ends up with an unconstrained model III 2HDM~\cite{dht}. The
Wilson coefficients of operators relevant to $b\rightarrow s
l^+l^-$ in the effective Hamiltonian have been calculated in the
model III 2HDM using the on-shell renormalization
prescription~\cite{et}. Some Feynman diagrams containing the
coupling of a neutral Higgs boson to the charged Higgs and
Goldstone bosons were missed in the calculations in ~\cite{et}. We
calculate them, including all the contributions, based on the
$\overline{MS}$ renormalization scheme. We then analyze the CP
asymmetry in $B_{d,s} \to l^+ l^-$ (l=$\mu, \tau$) using the
Wilson coefficients obtained by us. It is shown that the CP
asymmetry can be as large as  $40\%$ for $B^0_d$ and $4\%$ for
$B^0_s$ in the reasonable region of parameters in the model.

The organization of the paper is as follows. In Section II we
describe the model III 2HDM briefly. In section III we  give the
effective Hamiltonian responsible for $b\rightarrow s l^+l^-$ in
the model. We present the formula for CP asymmetry in $B_{d,s} \to
l^+ l^-$ in Section IV. The Section V is devoted to numerical
results. In Section VI we draw conclusions and discussions.
Finally we give contributions to Wilson coefficients $C_{10},
C_{Q_i}$ from individual diagrams in the model III 2HDM in the
appendix.

\section{\bf The model III two Higgs doublet model}
In general one can have a Yukawa Lagrangian of the form

\be {\cal L}_{Y}= \eta^{U}_{ij} \bar Q_{i,L} \tilde H_1 U_{j,R} +
\eta^D_{ij} \bar Q_{i,L} H_1 D_{j,R} + \xi^{U}_{ij} \bar
Q_{i,L}\tilde H_2 U_{j,R} +\xi^D_{ij}\bar Q_{i,L} H_2 D_{j,R}
\,+\, h.c. \label{lyukmod3} \ee

\noindent where $H_i$, for i=1,2, are the two scalar doublets of a
2HDM, while $\eta^{U,D}$ and $\xi^{U,D}$ are the non-diagonal
matrices of the Yukawa couplings.

For convenience we can choose to express $H_1$ and $H_2$ in a
suitable basis such that only the $\eta_{ij}^{U,D}$ couplings
generate the fermion masses, i.e.\ such that
\be \langle H_1\rangle=\left(\ba{c} 0\\
 {v\over \sqrt{2} } \ea\right)
\,\,\,\, , \,\,\,\, \langle H_2\rangle=0 \,\,\,. \ee

\noindent The two doublets in the basis are of the form

\be \label{base} H_1=\frac{1}{\sqrt{2}}\left[\left(\ba{c} 0 \\
v+\phi^0_1 \ea\right)+ \left(\ba{c} \sqrt{2}\, G^+\\
i G^0\ea\right)\right]\,\,\,\,;\,\,\,\,
H_2=\frac{1}{\sqrt{2}}\left(\ba{c}\sqrt{2}\,H^+\\ \phi^0_2+i
A^0\ea\right) \,\,\,. \ee

\noindent where $G^{0,\pm}$ are the Goldstone bosons, $H^{\pm}$
and $A^0$ are the physical charged-Higgs boson and CP-odd neutral
Higgs boson respectively. The advantage of using the basis is that
the first doublet $H_1$ corresponds to the scalar doublet of the
SM while the new Higgs fields arise from the second doublet $H_2$.
So  $\phi^0_2$ do not have couplings to the gauge bosons of the
form $\phi^0_2ZZ$ or $\phi^0_2W^+W^-$.

In eq. (\ref{base}) $\phi^0_1$, $\phi^0_2$ are not the neutral mass
eigenstates but linear combinations of the CP-even neutral Higgs
boson mass eigenstates, $H^0$ and $h^0$: \bea \label{masseigen}
H^0 & = & \phi^0_1 \cos\alpha + \phi^0_2\sin\alpha \\
\nonumber h^0 & = & -\phi^0_1\sin\alpha + \phi^0_2 \cos\alpha
 \eea
\noindent where $\alpha$ is the mixing angle, such that for
$\alpha\!=\!0$, ($\phi^0_1$, $\phi^0_2$) coincide with the mass
eigenstates.

After diagonalizing the mass matrix of the quark fields, the
flavor changing (FC) part of the Yukawa Lagrangian becomes

\be {\cal L}_{Y,FC} = \hat\xi^{U}_{ij} \bar Q_{i,L}\tilde H_2
U_{j,R}
 +\hat\xi^D_{ij}\bar Q_{i,L} H_2 D_{j,R} \,+\, h.c. \label{lyukfc}
\ee

\noindent where $Q_{i,L}$, $U_{j,R}$, and $D_{j,R}$ now denote the
quark fields which are in mass eigenstates and \be
\hat\xi^{U,D}=(V_L^{U,D})^{-1}\cdot \xi^{U,D} \cdot V_R^{U,D}
\label{neutral}\,. \ee In eq. (\ref{neutral}) $V_{L,R}^{U,D}$ are
the rotation matrices acting on the up- and down-type quarks, with
left or right chirality respectively, so that
$V_{CKM}=(V_L^U)^{\dag}V_L^D$ is usual Cabibbo-Kobayashi-Maskawa
(CKM) matrix. Feynman rules of Yukawa couplings follows from eq.
(\ref{lyukfc}) and can be found in, e.g., ref.~\cite{ars,bck}. The
flavor changing neutral current (FCNC) couplings are given by the
matrices $\hat\xi^{U,D}$ and the charged FC couplings are given by
\bea \hat\xi^{U}_{\rm charged}\!&=&\!\hat\xi^{U}\cdot
V_{CKM}\nonumber\\
\hat\xi^{D}_{\rm charged}\!&=&\!V_{CKM}\cdot \hat\xi^{D}
\label{charged}\,. \eea Because the definition of the
$\xi^{U,D}_{ij}$ couplings is arbitrary, we can take the rotated
couplings as the original ones and shall write $\xi^{U,D}$ in
stead of $\hat{\xi}^{U,D}$ hereafter. It is worth to note that the
peculiar form of charged FC couplings, eq. (\ref{charged}), is of
an important distinction from popular model I and II and has
significant phenomenological effects different from those in
models I and II~\cite{xlc}.

In order to proceed we assume the Cheng-Sher ansatz~\cite{cs}
 \be
\xi^{U,D}_{ij}=\lambda_{ij}\,\frac{\sqrt{m_i m_j}}{v} \label{sher}
\ee \noindent which ensures that the FCNC within the first two
generations are naturally suppressed by small quark masses. In the
ansatz the residual degree of arbitrariness of the FC couplings is
expressed through the $\lambda_{ij}$ parameters which are of order
one and need to be constrained by the available experiments. In
the paper we choose $\xi^{U,D}$ to be diagonal for the sake of
simplicity so that besides Higgs boson masses only $\lambda_{tt}$
and $\lambda_{bb}$ in the quark sector and $\lambda_{ll}$ in the
lepton sector are the new parameters which enter into the Wilson
coefficients relevant to the rare leptonic B decays.

The Yukawa Lagrangian (\ref{lyukmod3}) has no discrete symmetry.
In general, the Higgs potential in 2HDM which has no discrete
symmetry may be CP conserved or CP violated~\cite{m3,bs}. In order
to concentrate on the effects of FC Yukawa couplings we assume the
potential is CP conserved. Therefore, the new source of CP
violation is only from the Yukawa couplings. In a general basis
$\Phi_i$ (i=1,2) the general CP conserved potential is given by:
\begin{eqnarray}
\label{poten} V(\Phi_1,\Phi_2) &=& \lambda_1 (\Phi_1^{\dag}
\Phi_1- v_1^2)^2
 +\lambda_2  (\Phi_2^{\dag} \Phi_2 -v_2^2)^2 \nonumber\\
&&+ \lambda_3 [ (\Phi_1^{\dag} \Phi_1 -v_1^2)+ (\Phi_2^{\dag}
\Phi_2- v_2^2) ]^2
\nonumber\\
&&+ \lambda_4 [ (\Phi_1^{\dag} \Phi_1) (\Phi_2^{\dag} \Phi_2)-
(\Phi_1^{\dag} \Phi_2) (\Phi_2^{\dag} \Phi_1) ]
\nonumber \\
&& + \lambda_5
 [ \mbox{Re}(\Phi_1^{\dag} \Phi_2)- v_1 v_2 ]^2
 + \lambda_6
 [ \mbox{Im}(\Phi_1^{\dag} \Phi_2)]^2 \nonumber \\
 && + \lambda_7 (\Phi_1^{\dag}\Phi_1- v_1^2)(Re (\Phi_1^{\dag}
 \Phi_2)-v_1v_2)\nonumber \\ && +
 \lambda_8 (\Phi_2^{\dag}\Phi_2- v_2^2)(Re (\Phi_1^{\dag} \Phi_2)-v_1v_2)\, .
 \label{eq2}
\end{eqnarray}
in which all the coupling constants $\lambda_i$ are real due to
hermiticity. The minimum of the potential is at
\begin{eqnarray}
<\Phi_1>=\left( \begin{array}{c}
0 \\
v_1
\end{array}
\right), \ \ \ \ <\Phi_2>=\left( \begin{array}{c}
0 \\
v_2
\end{array}
\right),
\end{eqnarray}
thus breaks $SU(2)\times U(1)$ down to $U(1)_{EM}$. For the sake
of simplicity and to decrease the number of the parameters we
assume $\lambda_1=\lambda_2$ and $\lambda_7=\lambda_8= 2\lambda_3+
\lambda_5/2$ in the potential (\ref{eq2}). Then we have seven
parameters altogether in the Higgs sector. One of them,
$v=\sqrt{v^2_1+v^2_2}$, is fixed by the $W$ boson mass, $M^2_W =
\frac{1}{2}g^2 v^2$. The others can be expressed in terms of
mixing angles $\alpha^{\prime}$ and $\beta$ and Higgs masses
$m_{H^{\pm}}, m_{A^0}, m_{H^0}, m_{h^0}$. From the potential, it
is straightforward to derive \bea \label{ghgh}
g_{A^0H^-G^+}&=&\frac{g(m^2_{A^0}-m^2_{H^{\pm}})}{2 m_W}\; ,\\
\nonumber g_{H^0H^-G^+}&=&\frac{i g
\sin(\beta-\alpha^{\prime})(m^2_{H^0}-m^2_{H{\pm}})}{2 m_W}\; ,\\
\nonumber g_{h^0H^-G^+}&=&\frac{- i g
\cos(\beta-\alpha^{\prime})(m^2_{h^0}-m^2_{H{\pm}})}{2 m_W}\;
,\eea which have the same forms as those in a general model II
2HDM~\cite{2hdm,csb}.

The specific basis ($H_1, H_2$), eq. (\ref{base}), can be obtained
by a unitary transformation \be \label{ut}
H=S\Phi,\;\;\;\\
S=  \left(
\begin{array}{cc}
\cos\beta & \sin\beta \\
-\sin\beta & \cos\beta
\end{array}
\right),
 \ee
 where $H=(H_1, H_2)^T$, $\Phi=(\Phi_1, \Phi_2)^T$, and
 $\tan\beta=v_2/v_1$.

\section{\bf Effective Hamiltonian for $b\rightarrow s(d) l^+ l^-$}

The effective Hamiltonian describing the flavor changing processes
$b\rightarrow s(d) l^+ l^-$ can be defined as \be \label{halm}
H_{eff} = -\frac{4 G_F}{\sqrt{2}} \lambda_t
(\sum_{i=1}^{10}C_i(\mu)O_i(\mu)
+\sum_{i=1}^{10}C_{Q_i}(\mu)Q_i(\mu)). \ee where $\lambda_t=V_{tb}
V_{ts}^*$, $O_i^{\prime}s (i=1,\cdots,10)$ are the same as those
given in the ref.\cite{buras,wise}, and $Q_i^{\prime}$s come from
exchanging neutral Higgs bosons and have been given in refs.
~\cite{dhh,hly}.

 The explicit expressions of relevant operators in the Model III 2HDM for the process
 $B_{q} \rightarrow  l^+ l^-$ (q=d,s) are
 \begin{eqnarray*}
 O_{10} &=& \frac{e^2}{16\pi^2}(\bar{q}_L^\alpha \gamma_\mu
 b_L^\alpha)(\bar{l}\gamma_\mu \gamma_5 l) \\
 Q_1 &=& \frac{e^2}{16\pi^2}(\bar{q}_L^\alpha b_R^\alpha)(\bar{l}l) \\
 Q_2 &=& \frac{e^2}{16\pi^2}(\bar{q}_L^\alpha b_R^\alpha)(\bar{l}\gamma_5 l)
 \end{eqnarray*}

We calculate their Wilson coefficients at one loop level in the
$\overline{MS}$ renormalization scheme and results are given in
Appendix. Before proceeding, a remark is in place. In calculations
of the Wilson coefficients $C_{Q_i}$ the neutral Higgs boson $h^0$
($H^0$, $A^0$)-penguin diagrams with the charged Higgs  and
Goldstone bosons in the loop, which is missed in ref.~\cite{et},
should be included. In the Appendix we write the contributions of
these diagrams to $C_{Q_i}$ in terms of the general coupling
constants $g_{A^0H^-G^+}$, $g_{H^0H^-G^+}$ and $g_{h^0H^-G^+}$.
However, in numerical calculations we shall confine ourself to the
specific case of the couplings (\ref{ghgh}).

The QCD corrections to coefficients $C_i$ and $C_{Q_i}$ can be
incorporated in the standard way by using the renormalization
group equations. $Q_i(i=1,\cdots,10)$ does not mix with $O_{10}$
so that the evolution of $C_{10}$ remains unchanged and is given
by~\cite{wise}
\begin{eqnarray}
C_{10}(m_b)&=&C_{10}(m_W).
\end{eqnarray}

It is obvious that operators $O_i(i=1,\cdots,10)$ and
$Q_i(i=3,\cdots,10)$ do not mix into $Q_1$ and $Q_2$ and also
there is no mixing between $Q_1$ and $Q_2$. Therefore, the
evolution of $C_{Q_1},C_{Q_2}$ is controlled by the anomalous
dimensions of $Q_1,Q_2$ respectively~\cite{dhh}, i.e.,
\begin{eqnarray}
C_{Q_i}(m_b)&=&\eta^{-\gamma_Q/\beta_0}C_{Q_i}(m_W), \nnb\\
&=&1.24 \;\; C_{Q_i}(m_W),~~i=1,2,
\end{eqnarray}
where $\gamma_Q=-4$ \cite{h} is the one loop anomalous dimension
of $\bar{q}_Lb_R$, $\eta =
\frac{\alpha_s(m_b)}{\alpha_s(M_W)}\approx 1.72$, and $\beta_0 =
11 - (2/3)n_f = 23/3$.

\section{\bf CP asymmetry in $B_{d,s}\rightarrow l^+l^-$}

We need to know what kind of CP violating observables can be
defined in the process $B_q\rightarrow l^+l^-$ for q=d, s. At
first, direct CP violation is absent in this process due to
absence of strong phases. T-odd projection of polarization is a
kind of useful tool to probe the CP violating effects, for
example, in $\bxll$ \cite{ks,hl,hz}. However for the process we
are discussing here, we have actually only one independent
momentum and one independent spin which can be chosen as those of
$l^-$, so no T-odd projections can be defined. Unlike the case
generally discussed for hadronic final states, for example, that
in Ref.~\cite{gron}, the detected final states of $l^+$ and $l^-$
of this process in experiments are basically two asymptotic
energy-momentum eigenstates which are not CP eigenstates.
Considering for instance $B^0$ decays to $l^+ l^-$ in the rest
frame of $B^0$, due to the energy-momentum conservation we denote
the four-momenta of $l^-$ and $l^+$ as $p=(E,\vec{p})$ and
$\bar{p}=(E,-\vec{p})$. Then the angular momentum conservation
tells us that $l^+_L l^-_R$ and $l^+_R l^-_L$ final states are
forbidden. Hence we are left with a pair of CP conjugated final
states, $l^+_L l^-_L$ and $l^+_R l^-_R$, and the couple of the
corresponding CP conjugated process. Therefore, we may define the
time dependent CP asymmetries as~\cite{hl1} \beq A^1_{CP}(t)&=&
\frac{\Gamma(B^0_{phys}(t) \rightarrow l^+_L l^-_L)
-\Gamma(\bar{B}^0_{phys}(t) \rightarrow l^+_R l^-_R)}
{\Gamma(B^0_{phys}(t) \rightarrow l^+_L l^-_L)
+\Gamma(\bar{B}^0_{phys}(t) \rightarrow l^+_R l^-_R)}
\label{cpt1} \\
A^2_{CP}(t)&=& \frac{~\Gamma(B^0_{phys}(t) \rightarrow l^+_R
l^-_R) -\Gamma(\bar{B}^0_{phys}(t) \rightarrow l^+_L l^-_L)}
{\Gamma(B^0_{phys}(t) \rightarrow l^+_R l^-_R)
+\Gamma(\bar{B}^0_{phys}(t) \rightarrow l^+_L l^-_L)} \label{cpt2}
\eeq Two corresponding time integrated CP asymmetries are \beq
A^i_{CP}&=& \frac{\int^\infty_0 dt ~\Gamma(B^0_{phys}(t)
\rightarrow f_i) -\int^\infty_0 dt ~\Gamma(\bar{B}^0_{phys}(t)
\rightarrow \bar{f}_i)} {\int^\infty_0 dt ~\Gamma(B^0_{phys}(t)
\rightarrow f_i) +\int^\infty_0 dt ~\Gamma(\bar{B}^0_{phys}(t)
\rightarrow \bar{f}_i)}~~~~~~~i=1, 2 \label{cp1} \eeq Where
$f_{1,2}=l^+_{L,R}l^-_{L,R}$ with $l_{L(R)}$ being the helicity
eigenstate of eigenvalue $-1 (+1)$, $\bar{f}$ is the CP conjugated
state of $f$.

The time evolutions of the initial pure $B^0$ and $\bar{B}^0$
states are given by\cite{75} \beq
&& \big| B^0_{phys}(t) \big> = g_+(t)\aB0 + \frac{q}{p} g_-(t) \bB0, \nnb \\
&& \big| \bar{B}^0_{phys}(t) \big> = \frac{p}{q} g_-(t)\aB0 + g_+(t)\bB0. \label{b0t}
\eeq
with $g_{\pm}(t)$ given by
\beq
&& g_+(t) = exp(-\frac{1}{2}\Gamma t-i m t) cos(\frac{\Delta m}{2} t), \nnb \\
&& g_-(t) = exp(-\frac{1}{2}\Gamma t-i m t) i sin(\frac{\Delta m}{2} t) \label{gt}
\eeq

 The absence of strong phases implies
\beq |A_f|=|\bar{A}_{\bar{f}}|, ~~~~|A_{\bar{f}}|=|\bar{A}_f|
\label{axi} \eeq where $A_f(\bar{A}_f)= < f|  {\cal H}_{eff} | B^0
(\bar{B}^0)>$. And the CPT invariance leads to \beq
\frac{\bar{A}_f}{A_f} =  \bigg(
\frac{A_{\bar{f}}}{\bar{A}_{\bar{f}}} \bigg)^*\;. \label{rate11}
\eeq

In the above definition of CP asymmetry we need to separate the
final state $l^+_L l^-_L$ from $l^+_R l^-_R$  in order to measure
CP asymmetry. For l=$\tau$, the polarization analysis is
straightforward. However, detecting tau's is difficult
experimentally. For l=$\mu$, in principle one can separate the
final state $\mu^+_L \mu^-_L$ from $\mu^+_R \mu^-_R$ by measuring
the energy spectra of the electron from muon decay\cite{pes}. A
$\mu_L$ will decay to an energetic $e_L$, which must go forward to
carry the muon spin, and a less energetic pair of neutrino and
antineutrino because the electron is always
left-handed\footnote{In the present case it is quite a good
approximation to ignore the mass of electron.}{nofootinbib} and the
energy-momentum and angular momentum are conserved. Due to the
same reason, for  $\mu_R$, the relative energies of electron and a
pair of neutrino and antineutrino are roughly reversed. Therefore,
the energy spectra of the electron from the muon decay  is a
powerful $\mu$ spin analyzer. However, in practice muons never
decay in a 4$\pi$ detector because the lifetime of a muon is long
( c$\tau$=659 m). As pointed out in ref.\cite{hl1}, a possible way
to measure a polarized muon decay is to build special detectors
which can make muons lose its energy but keep polarization so that
the polarized muon decays can be measured.

In order to make measurements accessible, as proposed in ref.
\cite{hl2}, one can define the CP violating observable as
\bea &&A_{CP} = \frac{D}{S},\nnb\\
&& D =
\int^\infty_0 dt ~\sum_{i=1,2}\Gamma(B^0_{phys}(t) \rightarrow f_i) \nnb\\
&& -\int^\infty_0 dt ~\sum_{i=1,2}\Gamma(\bar{B}^0_{phys}(t) \rightarrow \bar{f}_i), \nnb\\
&& S = \int^\infty_0 dt ~\sum_{i=1,2}\Gamma(B^0_{phys}(t) \rightarrow f_i) \nnb\\
&& +\int^\infty_0 dt ~\sum_{i=1,2}\Gamma(\bar{B}^0_{phys}(t) \rightarrow \bar{f}_i)
\label{cp}
\eea
 Different from $A_{CP}^i$ (i=1,2), it
is accessible experimental to measure such defined CP asymmetry
$A_{CP}$. So in the following we shall concentrate on it.

 Using the effective Hamiltonian,
we obtain by a straightforward calculation \footnote {We have neglected the contributions, which is smaller
than or equal to $10^{-3}$ of the leading term, from the penguin diagrams with c and u quarks in the loop.
It is true for both $B_d$ and $B_s$ decays\cite{bur}. Therefore, although there are  weak phases from the c
or u quark in the loop, in particular,
 for $B_d$, the effect on the decay phase induced by them is negligibly small.}
\beq
\frac{\bar{A}_{f_1}}{A_{f_1}} =
 -\frac{\lambda_t}{\lambda_t^*} \frac{C_{Q1}\sqrt{1-4\hat{m}_l^2}
+ (C_{Q2}+2 \hat{m}_l C_{10})}{C^*_{Q1}\sqrt{1-4\hat{m}_l^2} -
(C^*_{Q2}+2 \hat{m}_l C^*_{10})},\label{rate} \eeq where
$\lambda_t=V_{tb} V_{td}^*$ or $V_{tb} V_{ts}^*$, $\hat{m}_l =
m_l/m_{B^0}$ and $C_i$'s are the Wilson coefficients at the
$m_b$ scale. 
Because $C_{Q_i}$'s are proportional to $m_l$ and $C_{10}$ is
independent of $m_l$ it follows from eq. (\ref{rate}) that the CP
asymmetry in $B_{d,s} \to l^+ l^-$ is independent of the mass of
the lepton in the approximation 0f neglecting $4\hat{m}_l^2$. That
is, it is the same for l = electron, muon.

In SM, one has\cite{nir}\footnote{Note that the phase convention
between $B^0$ and $\bar{B}^0$ is fixed as ${\cal CP}|B^0> = -
|\bar{B}^0>$ when deriving eqs. (\ref{rate}), (\ref{mix}).} \beq
\frac{q}{p}= - \frac{M^*_{12}}{|M_{12}|}= -
\frac{\lambda^*_t}{\lambda_t}, \label{mix} \eeq up to the
correction smaller than or equal to order of $10^{-2}$, $C_{10}$
is real, $C_{Q_2}=0$, and $C_{Q_1}$ is negligibly small. So it
follows from Eqs.(\ref{rate}), (\ref{mix}) that  there is no CP
violation in SM. \footnote{One can check by combining Eqs.
(\ref{mix}) and (\ref{rate}) that all freedoms of phase
conventions are cancelled out completely in
$\frac{q}{p}\frac{\bar{A}_{f_1}}{A_{f_1}}$, including the one
between $B^0$ and $\bar{B}^0$.  } If  one includes the correction
smaller than order of $10^{-2}$ to $|q/p|$=1\footnote{According to
the box diagram calculation in SM, the deviation of $|q/p|$ from 1
is $\sim 10^{-3} (10^{-5})$ for $B_d (B_s)$\cite{nir}. So
$10^{-2}$ is a conservative estimate.} one will have CP violation
of order of $10^{-3}$ for $B^0_d$ and $10^{-4}$ for $B^0_s$ which
are unobservably small.

In the approximation $|{q \over p}|=1$ the time integrated CP
asymmetry is \bea A_{CP} &=& - \frac{2 Im(\xi) X_q} {(1+
|\xi|^2)(1+X_q^2)}, ~~~q=d,s,\label{app} \eea where $X_q=
\frac{\Delta m_q}{\Gamma}$($q=d,s$ for $B^0_d$ and $B^0_s$
respectively),  and $\xi$ is \bea \xi &=&
 \frac{C_{Q1}\sqrt{1-4\hat{m}_l^2}
+ (C_{Q2}+2 \hat{m}_l C_{10})}{C^*_{Q1}\sqrt{1-4\hat{m}_l^2} -
(C^*_{Q2}+2 \hat{m}_l C^*_{10})}\label{rate1} \eea  As expected,
it is nonzero in the presence of CP violating phases. In deriving
Eq. (\ref{app}) we have used eq. (\ref{mix}) which is the result
for $B^0$-$\bar{B}^0$ mixing in the SM. $B^0$-$\bar{B}^0$ mixing
in the model II 2HDM has been examined in ref.~\cite{bcrs}. In the
model III there are new CP violating phases which might affect the
mixing. However, for the values of parameters which we assumed (
see below, eq.(\ref{lam})) the correction to the SM value of $q/p$
is below $20 \%$ and consequently  we can still use eq.
(\ref{app}) as an approximation. Because $X_s$ is larger than
$19.0$($90 \% ~CL$) and $X_d$ is just about $0.76$~\cite{data}
$A_{CP}$ in $B_s$ decays is much smaller than that in $B_d$
decays, as can be seen from eq. (\ref{app}).

\section{\bf Numerical results}

\noindent In numerical calculations, the following values of
parameters are assumed: \be \label{para}
 m_{A^0}=120 GeV,\;\;m_{h^0}=115 GeV,\;\; m_{H^0}=160 GeV,\;\; m_{H^\pm}=200 GeV\;. \ee

To study the dependence of $A_{CP}$ on masses of Higgs bosons, we
also include below some results obtained with the doubled values
of eq. (\ref{para}) for Higgs boson masses.

The plots of $|C_{Q_1}|$ and $|C_{Q_2}|$ for $\tau$ versus
$|\lambda_{bb}|$ for $|\lambda_{\tau\tau}|=5$ and fixed
$|\lambda_{tt}|$ are shown in Fig. 1 and Fig. 2 respectively. One
can see from the figures that $|C_{Q_i}|$ increases when
$|\lambda_{bb}|$ increases and the increase is faster for a larger
$|\lambda_{tt}|$ than that for a small $|\lambda_{tt}|$. Comparing
with the model II 2HDM, $|C_{Q_i}|$ can reach the values larger
than those in the model II if the value of $|\lambda_{bb}|$, taken
to be $>>1$, is equal to that of $\tan\beta$ in the model II and
the parameter $|\lambda_{tt}|$ is larger than $1/|\lambda_{bb}|$.
We calculate the constraint on $|\lambda_{bb}|$ and
$|\lambda_{tt}|$ for $|\lambda_{\mu\mu}|=50$ due to the
experimental upper bound of $Br(B_s\rightarrow \mu^+\mu^-)$ and
the result is shown in Fig. 3 where the horizontal line represents
the upper limit of $|\lambda_{tt}|$ which comes from the
experimental constraints of $B-\bar{B}$ mixing , $\Gamma(b \to s
\gamma)$, $\Gamma(b \to c \tau \bar{\nu}_\tau)$, $\rho_0$, $R_b$
and electric dipole moments (EDMs) of the electron and
neutron~\cite{bck}. The region under the curve (and the horizontal
line) is allowed by the experimental bound of Br($B_s\rightarrow
\mu^+\mu^-$).  We also calculate the corresponding constraint by
including the phases of the two parameters. The result is that the
constraint is not sensitive to the phase of $\lambda_{bb}$ or
$\lambda_{tt}$, as expected.

In numerically calculating $A_{CP}$, the following values of
parameters are assumed: \be\label{lam} |\lambda_{bb}|=50,\;\;
|\lambda_{tt}|=0.02, \ee \be\label{lam0} |\lambda_{\tau\tau}|=5,
50,\;\;|\lambda_{\mu\mu}|=5, 50,\;\;
\theta_{\tau\tau}=\pi/4,\;\;\theta_{\mu\mu}=\pi/4, \ee and
\be\label{lam1} \theta_{tt}+\theta_{bb}=\pi/2, \pi/3 \ee has been
set for simplicity. The values of Higgs boson masses and
$\lambda_{bb,tt}$, i.e., eqs. (\ref{para}), (\ref{lam}), and
(\ref{lam1}) have been set to satisfy the experimental constraints
of $B-\bar{B}$ mixing , $\Gamma(b \to s \gamma)$, $\Gamma(b \to c
\tau \bar{\nu}_\tau)$, $\rho_0$, $R_b$ and electric dipole moments
(EDMs) of the electron and neutron~\cite{bck}. The value of $
|\lambda_{\tau\tau}|$, eq. (\ref{lam0}), has been set to satisfy
the constraint from $Z\rightarrow l^+l^-$ which has been analyzed
and the upper limit is 56~\cite{iltan}. The Figs. 4 and 5 are
devoted to $A_{CP}$ versus $\theta_{tt}$ for $B^0_d\rightarrow
\tau^+\tau^-$ and $B^0_d\rightarrow \mu^+\mu^-$ respectively.
$A_{CP}$ depends on $\theta_{tt}$ significantly and can reach
$40\%$ and $35\%$ for $l=\tau$ and $l=\mu$ respectively. As for
the $B^0_s$ decays, $A_{CP}$ is much smaller than that in $B^0_d$
decays, e.g., it can only reach $4\%$ for $B^0_s\rightarrow
\tau^+\tau^-$, as expected. In order to see the effects of
different parameters we consider two cases (a) and (b) which
correspond $|\lambda_{\tau\tau}|$ (or $|\lambda_{\mu\mu}|$) =5 and
50 respectively and plot two curves in each case in Figs. 4,5. In
Fig. 4 the solid (dotted) one corresponds the set of Higgs boson
masses same as eq. (29) (doubled values of those in eq. (29)). One
can see from Fig. 4a that the maximum of $A_{CP}$ significantly
decreases when the masses of Higgs bosons increase. However, Fig.
4b shows that the change is small when the masses of Higgs bosons
are doubled. The reason is that in the case (a) $|C_{Q_i}|$
(i=1,2) for low Higgs masses, eq.(29), is the same order of
magnitude as $\hat{m}_{\tau} C_{10}$ and $C_{10}$ is almost real,
when Higgs boson masses increase $C_{Q_i}$ decreases but $C_{10}$
keeps unchanged so that the CP asymmetry decreases, as shown in
Fig. 4a. In the case (b), $C_{Q_i}$ is much larger than
$\hat{m}_{\tau} C_{10}$. Therefore, when Higgs boson masses
increase although $C_{Q_i}$ is decreased $\xi$ is almost unchanged
since the numerator and dominator are almost simultaneously scaled
(see, eq. (\ref{rate1}) ). The small change of $A_{CP}$ is due to
the increase of $m_{H^{\pm}}$. In order to see the effect of the
phase $\theta_{tt}+\theta_{bb}$ on $A_{CP}$ we plot two curves in
Fig. 5 where the solid (dotted) one corresponds
$\theta_{tt}+\theta_{bb}=\pi/2\; (\pi/3)$. One can see from the
figure that the curve moves toward the direction opposite to that
of the transverse axis when $\theta_{tt}+\theta_{bb}$ decreases.

With the branching ratios 
\begin{eqnarray}
    Br~ (B^0_{q} \to l^+ l^-)
    &=& \frac{G_F^2 \alpha_{EM}^2}{64 \pi^3}
    m_{B_{q}}^3 \tau_{B_{q}} f^2_{B_{q}}
     |\lambda_t|^2
    \sqrt{1 - 4 {\hat m_l^2}}
    \nonumber \\
    && \times
    \left[ \left(1 - 4 {\hat m_l^2}\right)
     C_{Q_1}^2
    + \left( C_{Q_2}
    +2 {\hat m_l} C_{10} \right)^2 \right],
\end{eqnarray}
where $\tau_{B_{q}}$ is the $B_{q}$ lifetime,
 we calculate
the events  $N_q$  needed for observing $A_{CP}$ at 1$\sigma$ in
the areas of parameter space in which $A_{CP}$ and the branching
ratios both have large values and all experimental constraints are
satisfied. For l=$\mu$, they are order of $10^8$ and $10^9$ for
$B_d^0$ and $B_s^0$ respectively. Therefore,
 $10^{10}$ ($10^{11}$) $B_d$ ($B_s$)
per year, which is in the designed range in the future B factors
with $10^8$- $10^{12}$ B hadrons per year~\cite{bs}, is needed in
order to observe the CP asymmetry in $B\rightarrow \mu^+\mu^-$
with good accuracy. For l=$\tau$, the events $N_q$ are order of
$10^6$ and $10^7$ for $B_d^0$ and $B_s^0$ respectively. Assuming a
total of $5\times 10^8 (10^9)$ $B_d\bar{B_d}$ ($B_s\bar{B_s}$)
decays, one can expect to observe $\sim 100 $ identified
$B_q\rightarrow \tau^+\tau^-$ events, permitting a test of the
predicted CP asymmetry with good accuracy.

\section{\bf Conclusions}

In summary, we have calculated the Wilson coefficients $C_{10},
C_{Q_i}$ (i=1,2) in the $\overline{MS}$ renormalization scheme in
the model III 2HDM. Comparing with the model II 2HDM, $|C_{Q_i}|$
can reach the values larger than those in the model II when the
value of $|\lambda_{bb}|$, taken to be $>>1$, is equal to that of
$\tan\beta$ in the model II and the parameter $|\lambda_{tt}|$ is
larger than $1/|\lambda_{bb}|$. It is shown that there is a
constraint on $\lambda_{bb}$ and $\lambda_{tt}$ due to the
experimental limit of Br($B_s\rightarrow \mu^+\mu^-$).

 We have analyzed the CP violation
in decays $B^0_q\rightarrow l^+l^-$ (q=d,s).  The CP asymmetry
depends on the parameters of models, in particular, the phase
$\theta_{tt}$ significantly. $A_{CP}$ in $B_d\rightarrow l^+l^-$
can be as large as $40\%$ and $35\%$ for $l=\tau$ and $l=\mu$
respectively. It can reach $4\%$ for $B^0_s$ decays. Because in
SM CP violation is smaller than or equal to O($10^{-3}$) which is
unobservably small, an observation of CP asymmetry in the decays
$B^0_q \to l^+l^- (q=d,s)$ would unambiguously signal the
existence of new physics.

\section*{Acknowledgments}
The work was supported in part by the National Nature Science
Foundation of China.

\section*{\bf Appendix}
By computing the self-energy type, Higgs-penguin and box diagrams,
$C^a_{Q_1}$ and $C^a_{Q_2}$ for $l=\tau$ with the superscript
denoting the type of a diagram or the contribution of exchanging a
specific Higgs boson are extracted out and given below
\begin{eqnarray}
 C_{Q_1}^{A^0} &=& \frac{x_t m_b m_\tau
 Im\lambda_{\tau\tau}i}{4 m_{A^0}^2 \sin^2\theta_W}
 \left\{\int_0^1 dx \int_0^{1-x} dy \left[
 \frac{y_t \lambda_{tt}^\ast(2-x)}{S_1(y_t,x_t)}+\frac{y_t \lambda_{tt}(1+x)}{S_1(x_t,y_t)}
 \right. \right. \nonumber \\
 &+& \left. \left.
 |\lambda_{tt}|^2\lambda_{bb}\left(1+2\ln{\frac{S_2(y_t)m_{H^-}^2}{\mu^2}}
 +\frac{y_t (\lambda_{tt}-2i Im\lambda_{tt} x)}{S_2(y_t) \lambda_{bb}}
 -\frac{y_t \lambda_{tt}}{S_2(y_t) \lambda_{tt}^\ast} \right) \right. \right.\
 \nonumber \\
 &+& \left. \left. \lambda_{tt}^\ast \left(1+2\ln{\frac{S_2(x_t)m_W^2}{\mu^2}} \right)
 +\frac{\lambda_{bb}}{x_t}\left(1+2\ln{\frac{m_{H^-}^2
 S_1(y_t,x_t)}{x_t \mu^2}}\right) \right.\right. \nonumber \\
 &+& \left. \left.
 \frac{2 y_t g_{A^0 H^- G^+} }{g m_W}\left(\frac{\lambda_{bb}-\lambda_{tt}^\ast x}
 {S_1 (y_t,x_t)} + \frac{\lambda_{tt}(1-x)}{S_1 (x_t,y_t)}\right)
 -\frac{2[\lambda_{tt}^\ast+i Im\lambda_{tt}(2+x_t)x]}{S_2(x_t)}\right] \right. \nonumber \\
 &+& \left.
  \lambda_{tt}\lambda_{bb}^2 \left[\ln\frac{\mu^2}{m_{H^-}^2}-\int_0^1 dx \ln{S_4(y_t)}\right]
 +\lambda_{bb} \left[\ln\frac{\mu^2}{m_W^2}-\int_0^1 dx \ln{S_4(x_t)}\right]
 \right\}
 \end{eqnarray}

 \begin{eqnarray}
 C_{Q_2}^{A^0} &=& \frac{x_t m_b m_\tau
 Re\lambda_{\tau\tau}}{4 m_{A^0}^2 \sin^2\theta_W}
 \left\{\int_0^1 dx \int_0^{1-x} dy \left[
 \frac{y_t \lambda_{tt}^\ast(2-x)}{S_1(y_t,x_t)}+\frac{y_t \lambda_{tt}(1+x)}{S_1(x_t,y_t)}
 \right. \right. \nonumber \\
 &+& \left. \left.
 |\lambda_{tt}|^2\lambda_{bb}\left(1+2\ln{\frac{S_2(y_t)m_{H^-}^2}{\mu^2}}
 +\frac{y_t (\lambda_{tt}-2i Im\lambda_{tt} x)}{S_2(y_t) \lambda_{bb}}
 -\frac{y_t \lambda_{tt}}{S_2(y_t) \lambda_{tt}^\ast} \right) \right. \right.\
 \nonumber \\
 &+& \left. \left. \lambda_{tt}^\ast \left(1+2\ln{\frac{S_2(x_t)m_W^2}{\mu^2}} \right)
 +\frac{\lambda_{bb}}{x_t}\left(1+2\ln{\frac{m_{H^-}^2
 S_1(y_t,x_t)}{x_t \mu^2}}\right) \right.\right. \nonumber \\
 &+& \left. \left.
 \frac{2 y_t g_{A^0 H^- G^+} }{g m_W}\left(\frac{\lambda_{bb}-\lambda_{tt}^\ast x}
 {S_1 (y_t,x_t)} + \frac{\lambda_{tt}(1-x)}{S_1 (x_t,y_t)}\right)
 -\frac{2[\lambda_{tt}^\ast+i Im\lambda_{tt}(2+x_t)x]}{S_2(x_t)}\right] \right. \nonumber \\
 &+& \left.
  \lambda_{tt}\lambda_{bb}^2 \left[\ln\frac{\mu^2}{m_{H^-}^2}-\int_0^1 dx \ln{S_4(y_t)}\right]
 +\lambda_{bb} \left[\ln\frac{\mu^2}{m_W^2}-\int_0^1 dx \ln{S_4(x_t)}\right]
 \right\}
 \end{eqnarray}

 \begin{eqnarray}
 C_{Q_1}^{H^0} &=& \frac{x_t m_b m_\tau (Re\lambda_{\tau\tau}
 S_\alpha+C_\alpha)}{4m_{H^0}^2 \sin^2\theta_W}
 \left\{\int_0^1 dx \int_0^{1-x} dy
 \left[\frac{2(S_\alpha(\lambda_{tt}^\ast -2Re\lambda_{tt}
 x)+C_\alpha(1-2x))}{S_2(x_t)} \right. \right. \nonumber \\
 &+& \left. \left.
 \frac{S_\alpha y_t \lambda_{tt}^\ast(x-2)}{S_1(y_t,x_t)}
 -\frac{2x_t (Re\lambda_{tt} S_\alpha+C_\alpha)x}{S_2(x_t)}
 -\frac{y_t \lambda_{tt}
 S_\alpha (1+x)}{S_1(x_t,y_t)}
 +\frac{C_\alpha (4x-3x_t)}{x_t S_3(x_t)} \right. \right.
 \nonumber \\
 &-& \left. \left.
 (\lambda_{tt}^\ast S_\alpha +C_\alpha)\left(1+2\ln\frac{m_W^2
 S_2(x_t)}{\mu^2}\right)-\frac{S_\alpha \lambda_{bb}}{x_t}
 \left(1+2\ln{\frac{m_{H^-}^2 S_1(y_t,x_t)}{x_t \mu^2}}\right)
 \right. \right. \nonumber \\
 &-& \left. \left.
 \frac{C_\alpha}{x_t}\left(1+2\ln{\frac{m_W^2
 S_3(x_t)}{\mu^2}}\right)
 +|\lambda_{tt}|^2 \left(\frac{y_t (S_\alpha (\lambda_{tt}
 -2Re\lambda_{tt} x)+C_\alpha(1-2x))}{S_2(y_t)}
 \right. \right. \right. \nonumber \\
 &-& \left. \left. \left.
 \frac{y_t \lambda_{bb}(\lambda_{tt}S_\alpha +C_\alpha)}
 {\lambda_{tt}^\ast S_2(y_t)}-\frac{\lambda_{bb}(\lambda_{tt}^\ast S_\alpha +C_\alpha)}
 {\lambda_{tt}^\ast}\left(1+2\ln{\frac{m_{H^-}^2
 S_2(y_t)}{\mu^2}}\right) \right)
 \right.\right. \nonumber \\
 &+& \left.\left.
 \frac{2y_t g_{H^0 H^- G^+}}{i g m_W}\left(\frac{\lambda_{bb}-\lambda_{tt}^\ast x}
 {S_1 (y_t,x_t)} + \frac{\lambda_{tt}(1-x)}{S_1 (x_t,y_t)}\right)\right]
 \right. \nonumber\\
 &+& \left.
 \lambda_{tt} \lambda_{bb} (\lambda_{bb} S_\alpha
 +C_\alpha)\left[\int_0^1 dx \ln S_4(y_t) -\ln{\frac{\mu^2}{m_{H^-}^2}} \right]
 \right. \nonumber \\
 &+& \left. (\lambda_{bb} S_\alpha +C_\alpha)\left[\int_0^1 dx \ln S_4(x_t)
 -\ln{\frac{\mu^2}{m_W^2}} \right] \right\}
 \end{eqnarray}

 \begin{eqnarray}
 C_{Q_2}^{H^0} &=& \frac{x_t m_b m_\tau Im\lambda_{\tau\tau}
 S_\alpha i}{4m_{H^0}^2 \sin^2\theta_W}
 \left\{\int_0^1 dx \int_0^{1-x} dy
 \left[\frac{2(S_\alpha(\lambda_{tt}^\ast -2Re\lambda_{tt}
 x)+C_\alpha(1-2x))}{S_2(x_t)} \right. \right. \nonumber \\
 &+& \left. \left.
 \frac{S_\alpha y_t \lambda_{tt}^\ast(x-2)}{S_1(y_t,x_t)}
 -\frac{2x_t (Re\lambda_{tt} S_\alpha+C_\alpha)x}{S_2(x_t)}
 -\frac{y_t \lambda_{tt}
 S_\alpha (1+x)}{S_1(x_t,y_t)}
 +\frac{C_\alpha (4x-3x_t)}{x_t S_3(x_t)} \right. \right.
 \nonumber \\
 &-& \left. \left.
 (\lambda_{tt}^\ast S_\alpha +C_\alpha)\left(1+2\ln\frac{m_W^2
 S_2(x_t)}{\mu^2}\right)-\frac{S_\alpha \lambda_{bb}}{x_t}
 \left(1+2\ln{\frac{m_{H^-}^2 S_1(y_t,x_t)}{x_t \mu^2}}\right)
 \right. \right. \nonumber \\
 &-& \left. \left.
 \frac{C_\alpha}{x_t}\left(1+2\ln{\frac{m_W^2
 S_3(x_t)}{\mu^2}}\right)
 +|\lambda_{tt}|^2 \left(\frac{y_t (S_\alpha (\lambda_{tt}
 -2Re\lambda_{tt} x)+C_\alpha(1-2x))}{S_2(y_t)}
 \right. \right. \right. \nonumber \\
 &-& \left. \left. \left.
 \frac{y_t \lambda_{bb}(\lambda_{tt}S_\alpha +C_\alpha)}
 {\lambda_{tt}^\ast S_2(y_t)}-\frac{\lambda_{bb}(\lambda_{tt}^\ast S_\alpha +C_\alpha)}
 {\lambda_{tt}^\ast}\left(1+2\ln{\frac{m_{H^-}^2
 S_2(y_t)}{\mu^2}}\right) \right)
 \right.\right. \nonumber \\
 &+& \left.\left.
 \frac{2y_t g_{H^0 H^- G^+}}{i g m_W}\left(\frac{\lambda_{bb}-\lambda_{tt}^\ast x}
 {S_1 (y_t,x_t)} + \frac{\lambda_{tt}(1-x)}{S_1 (x_t,y_t)}\right)\right]
 \right. \nonumber \\
 &+& \left.
 \lambda_{tt} \lambda_{bb} (\lambda_{bb} S_\alpha
 +C_\alpha)\left[\int_0^1 dx \ln S_4(y_t) -\ln{\frac{\mu^2}{m_{H^-}^2}} \right]
 \right. \nonumber \\
 &+& \left. (\lambda_{bb} S_\alpha +C_\alpha)\left[\int_0^1 dx \ln S_4(x_t)
 -\ln{\frac{\mu^2}{m_W^2}} \right] \right\}
 \end{eqnarray}

 \begin{eqnarray}
 C_{Q_1}^{h^0} &=& \frac{x_t m_b m_\tau (Re\lambda_{\tau\tau}
 C_\alpha-S_\alpha)}{4m_{h^0}^2 \sin^2\theta_W}
 \left\{\int_0^1 dx \int_0^{1-x} dy
 \left[\frac{2(C_\alpha(\lambda_{tt}^\ast -2Re\lambda_{tt}
 x)+S_\alpha(2x-1))}{S_2(x_t)} \right. \right. \nonumber \\
 &+& \left. \left.
 \frac{C_\alpha y_t \lambda_{tt}^\ast (x-2)}{S_1(y_t,x_t)}
 -\frac{2x_t (Re\lambda_{tt} C_\alpha -S_\alpha)x}{S_2(x_t)}
 -\frac{y_t \lambda_{tt} C_\alpha(1+x)}{S_1(x_t,y_t)}
 +\frac{S_\alpha (3x_t-4x)}{x_t S_3(x_t)} \right. \right.
 \nonumber \\
 &-& \left. \left.
 (\lambda_{tt}^\ast C_\alpha -S_\alpha)\left(1+2\ln\frac{m_W^2
 S_2(x_t)}{\mu^2}\right)-\frac{C_\alpha \lambda_{bb}}{x_t}
 \left(1+2\ln{\frac{m_{H^-}^2 S_1(y_t,x_t)}{x_t \mu^2}}\right)
 \right. \right. \nonumber \\
 &+& \left. \left.
 \frac{S_\alpha}{x_t}\left(1+2\ln{\frac{m_W^2
 S_3(x_t)}{\mu^2}}\right)
 +|\lambda_{tt}|^2 \left(\frac{y_t (C_\alpha (\lambda_{tt}
 -2Re\lambda_{tt} x)+S_\alpha(2x-1))}{S_2(y_t)}
 \right. \right. \right. \nonumber \\
 &-& \left. \left. \left.
 \frac{y_t \lambda_{bb}(\lambda_{tt}C_\alpha -S_\alpha)}
 {\lambda_{tt}^\ast S_2(y_t)}-\frac{\lambda_{bb}(\lambda_{tt}^\ast C_\alpha -S_\alpha)}
 {\lambda_{tt}^\ast}\left(1+2\ln{\frac{m_{H^-}^2
 S_2(y_t)}{\mu^2}}\right) \right)
 \right.\right. \nonumber\\
 &-& \left.\left.
 \frac{2 y_t g_{h^0 H^- G^+}}{i g m_W}\left(\frac{\lambda_{tt}^\ast x-\lambda_{bb}}
 {S_1 (y_t,x_t)} + \frac{\lambda_{tt}(x-1)}{S_1 (x_t,y_t)}\right) \right]
 \right. \nonumber \\
 &+& \left.
 \lambda_{tt} \lambda_{bb} (\lambda_{bb} C_\alpha
 -S_\alpha)\left[\int_0^1 dx \ln S_4(y_t) -\ln{\frac{\mu^2}{m_{H^-}^2}} \right]
 \right. \nonumber \\
 &+& \left. (\lambda_{bb} C_\alpha -S_\alpha)\left[\int_0^1 dx \ln S_4(x_t)
 -\ln{\frac{\mu^2}{m_W^2}} \right] \right\}
 \end{eqnarray}

 \begin{eqnarray}
 C_{Q_2}^{h^0} &=& \frac{x_t m_b m_\tau Im\lambda_{\tau\tau}
 C_\alpha i}{4m_{h^0}^2 \sin^2\theta_W}
 \left\{\int_0^1 dx \int_0^{1-x} dy
 \left[\frac{2(C_\alpha(\lambda_{tt}^\ast -2Re\lambda_{tt}
 x)+S_\alpha(2x-1))}{S_2(x_t)} \right. \right. \nonumber \\
 &+& \left. \left.
 \frac{C_\alpha y_t \lambda_{tt}^\ast (x-2)}{S_1(y_t,x_t)}
 -\frac{2x_t (Re\lambda_{tt} C_\alpha -S_\alpha)x}{S_2(x_t)}
 -\frac{y_t \lambda_{tt} C_\alpha(1+x)}{S_1(x_t,y_t)}
 +\frac{S_\alpha (3x_t-4x)}{x_t S_3(x_t)} \right. \right.
 \nonumber \\
 &-& \left. \left.
 (\lambda_{tt}^\ast C_\alpha -S_\alpha)\left(1+2\ln\frac{m_W^2
 S_2(x_t)}{\mu^2}\right)-\frac{C_\alpha \lambda_{bb}}{x_t}
 \left(1+2\ln{\frac{m_{H^-}^2 S_1(y_t,x_t)}{x_t \mu^2}}\right)
 \right. \right. \nonumber \\
 &+& \left. \left.
 \frac{S_\alpha}{x_t}\left(1+2\ln{\frac{m_W^2
 S_3(x_t)}{\mu^2}}\right)
 +|\lambda_{tt}|^2 \left(\frac{y_t (C_\alpha (\lambda_{tt}
 -2Re\lambda_{tt} x)+S_\alpha(2x-1))}{S_2(y_t)}
 \right. \right. \right. \nonumber \\
 &-& \left. \left. \left.
 \frac{y_t \lambda_{bb}(\lambda_{tt}C_\alpha -S_\alpha)}
 {\lambda_{tt}^\ast S_2(y_t)}-\frac{\lambda_{bb}(\lambda_{tt}^\ast C_\alpha -S_\alpha)}
 {\lambda_{tt}^\ast}\left(1+2\ln{\frac{m_{H^-}^2
 S_2(y_t)}{\mu^2}}\right) \right)
 \right.\right. \nonumber\\
 &-& \left.\left.
 \frac{2 y_t g_{h^0 H^- G^+}}{i g m_W}\left(\frac{\lambda_{tt}^\ast x-\lambda_{bb}}
 {S_1 (y_t,x_t)} + \frac{\lambda_{tt}(x-1)}{S_1 (x_t,y_t)}\right) \right]
 \right. \nonumber \\
 &+& \left.
 \lambda_{tt} \lambda_{bb} (\lambda_{bb} C_\alpha
 -S_\alpha)\left[\int_0^1 dx \ln S_4(y_t) -\ln{\frac{\mu^2}{m_{H^-}^2}} \right]
 \right. \nonumber \\
 &+& \left. (\lambda_{bb} C_\alpha -S_\alpha)\left[\int_0^1 dx \ln S_4(x_t)
 -\ln{\frac{\mu^2}{m_W^2}} \right] \right\}
 \end{eqnarray}

 \begin{equation}
 C_{Q_1}^{box}=\frac{x_t m_b m_\tau \lambda_{bb}
 \lambda_{\tau\tau}^\ast}{4m_{H^-}^2 \sin^2\theta_W}
 \int_0^1 dx \int_0^{1-x} dy \left[\frac{1}{S_1(y_t,x_t)}\right]
 \end{equation}

 \begin{equation}
 C_{Q_2}^{box}=\frac{-x_t m_b m_\tau \lambda_{bb}
 \lambda_{\tau\tau}^\ast}{4m_{H^-}^2 \sin^2\theta_W}
 \int_0^1 dx \int_0^{1-x} dy \left[\frac{1}{S_1(y_t,x_t)}\right]
 \end{equation}
Adding all the contributions to  $C_{Q_1}$ and  $C_{Q_2}$  given
above respectively, we get the Wilson coefficients $C_{Q_i}$
(i=1,2). In numerical calculations we use the couplings among the
neutral and charged Higgs bosons and charged Goldstone bosons
given in eq. (\ref{ghgh}).

The Wilson coefficient $C_{10}$ is
 \begin{eqnarray}
 C_{10}&=& C_{10}^{SM}+ C_{10}^{new},\nonumber \\
 C_{10}^{new} &=& \frac{x_t |\lambda_{tt}|^2} {48 \sin^2\theta_W}
 \left\{\int_0^1 dx \int_0^{1-x} dy \left[
 6(2\sin^2\theta_W-1)\ln\frac{m_{H^-}^2 S_3(y_t)}{\mu^2}
 \right. \right. \nonumber \\
 &-& \left. \left.
 \frac{2y_t(3-4\sin^2\theta_W)}{S_2(y_t)}
 -8\sin^2\theta_W \left(1+\ln\frac{m_{H^-}^2 S_2(y_t)}{\mu^2} \right) \right]
 \right. \nonumber\\
 &-& \left.
 (3-2\sin^2\theta_W)\left[2\int_0^1 dx (x-1)\ln S_4(y_t)
 +\ln{\frac{\mu^2}{m_{H^-}^2}} \right] \right\}
 \end{eqnarray}

 In above equations, the definitions of the functions $S_i$ are

 \begin{eqnarray}
 S_1(\rho,\phi) &=& \rho \phi (1-x-y) +x \phi +y \rho \nonumber \\
 S_2(\rho) &=& (1-x-y) + (x+y) \rho \nonumber \\
 S_3(\rho) &=& (1-x-y)\rho + (x+y)  \nonumber \\
 S_4(\rho) &=& 1-x+x\rho\;,
 \end{eqnarray}

 and

 \begin{equation}
 x_t=\frac{m_t^2}{m_W^2},\hspace{5mm}
 y_t=\frac{m_t^2}{m_{H^\pm}^2}\;,
 \end{equation}

 \begin{equation}
 C_\alpha=\cos \alpha, \hspace{5mm}
 S_\alpha=\sin \alpha,
 \end{equation}
 with $\alpha$ being the mixing angle of the CP-even neutral Higgs bosons.

\bibliographystyle{h-physrev}

\vspace{2cm}
\begin{figure}[b]
\begin{center}
\epsfig{file=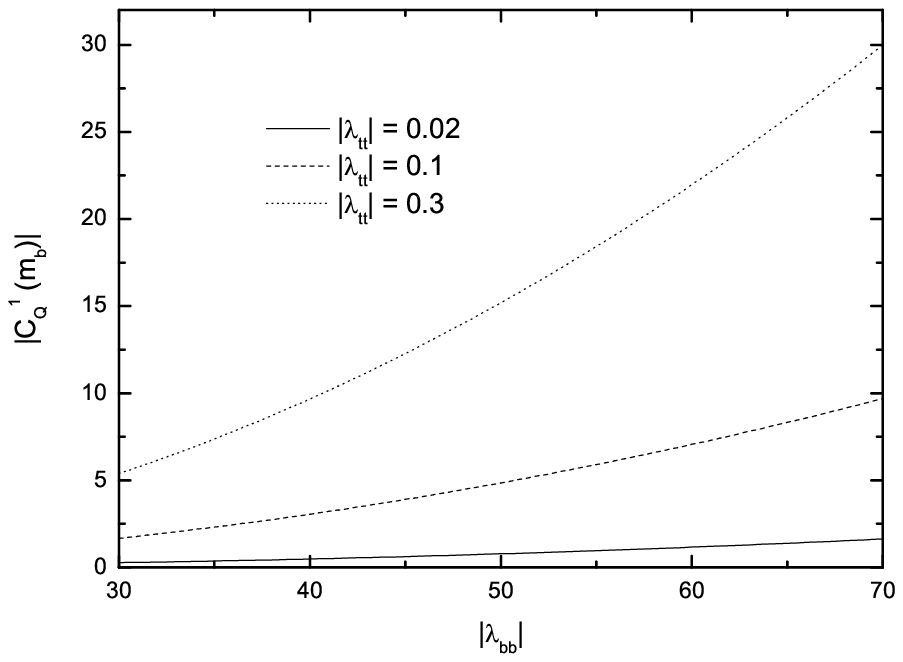,width=13cm} \vspace*{0.5cm}
\caption{$|C_{Q_1}|$ versus $|\lambda_{bb}|$. } \label{fig:cq1}
\end{center}
\end{figure}
\vspace*{-1cm}
\begin{figure}[b]
\begin{center}
\epsfig{file=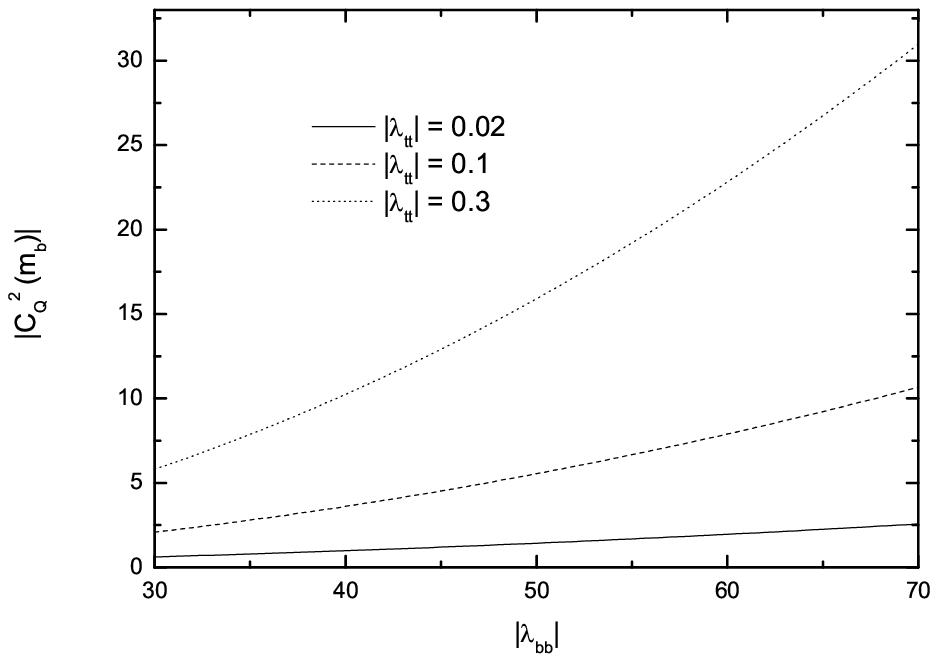,width=13cm} \vspace*{0.5cm}
\caption{$|C_{Q_2}|$ versus $|\lambda_{bb}|$. } \label{fig:cq2}
\end{center}
\end{figure}
\vspace*{-1cm}
\begin{figure}[b]
\begin{center}
\epsfig{file=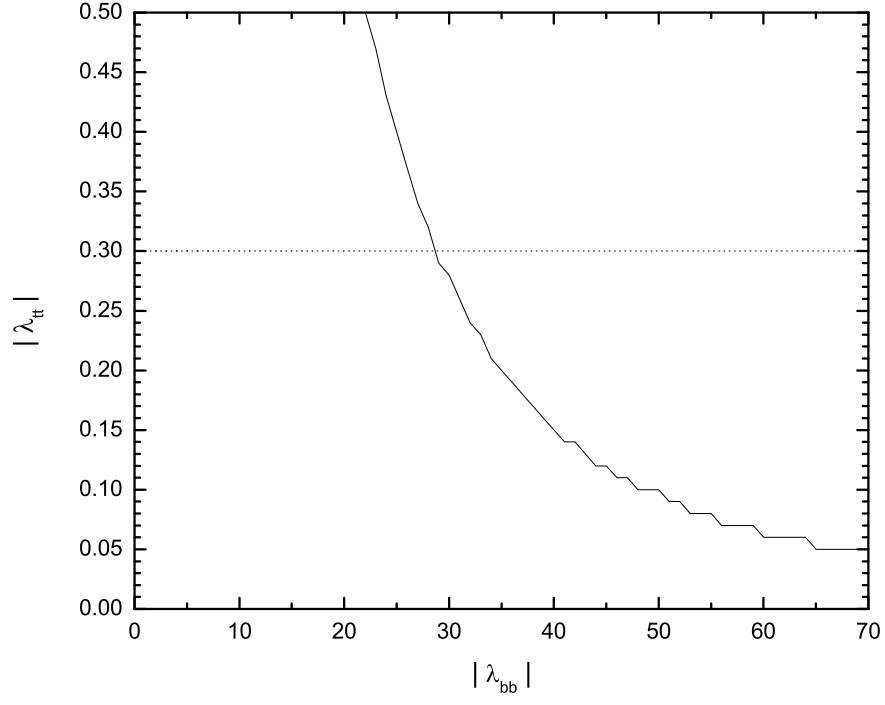,width=13cm} \vspace*{0.5cm} \caption{The
constraint on $|\lambda_{bb}|$ and $|\lambda_{tt}|$ due to the
experimental upper bound of $Br(B_s\rightarrow \mu^+\mu^-)$.}
\label{fig:bound}
\end{center}
\end{figure}
\vspace*{-1cm}
\begin{figure}[b]
\begin{center}
\begin{tabular}{cc}
\epsfig{file=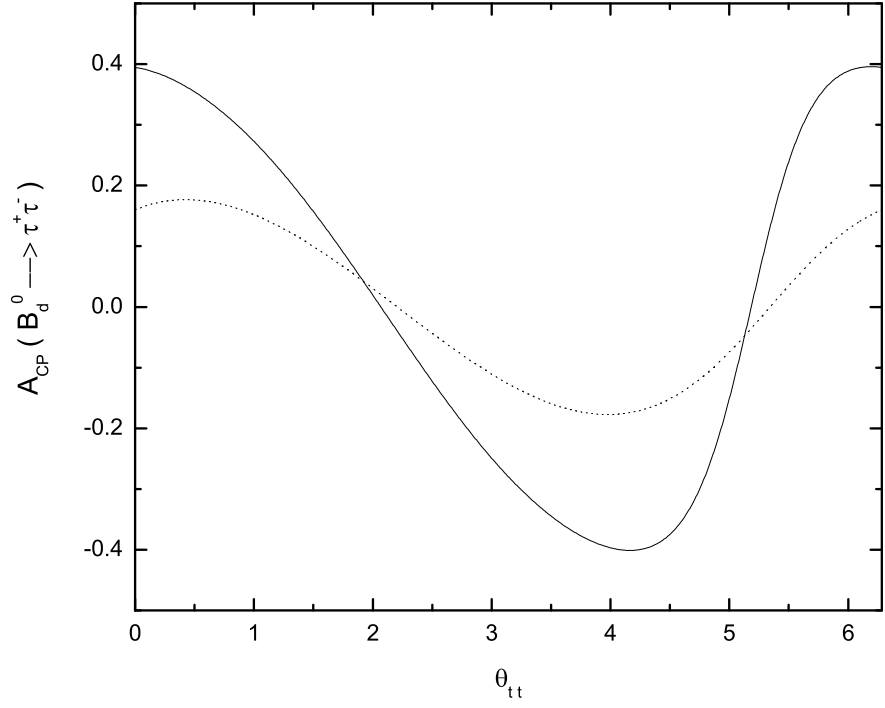,width=13cm}\\
(a)\\
\epsfig{file=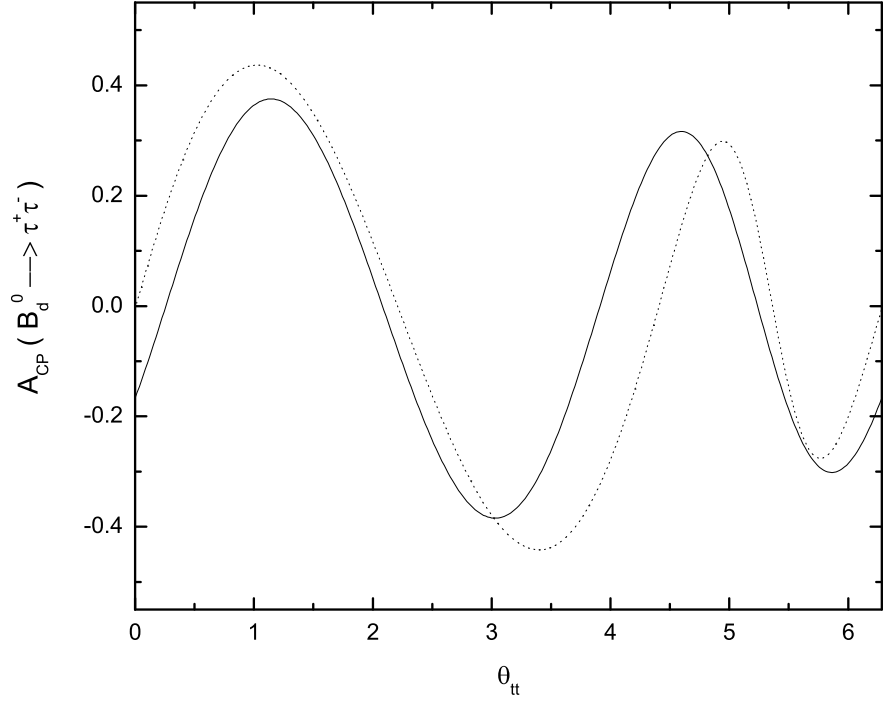,width=13cm}\\
(b)\\
\end{tabular}
\caption{(a) $A_{CP}$ for $B^0_d\rightarrow \tau^+\tau^-$ versus
the CP violating phase $\theta_{tt}$, for
$|\lambda_{\tau\tau}|=5$. The solid line stands for the masses of
Higgs bosons same as eq. (29), the dotted line stands for the masses of Higgs bosons same as doubled those in eq. (29). (b)The same as (a), except
for $|\lambda_{\tau\tau}|=50$.} \label{fig:acpd}
\end{center}
\end{figure}
\vspace*{3cm}
\begin{figure}[b]
\begin{center}
\begin{tabular}{cc}
\epsfig{file=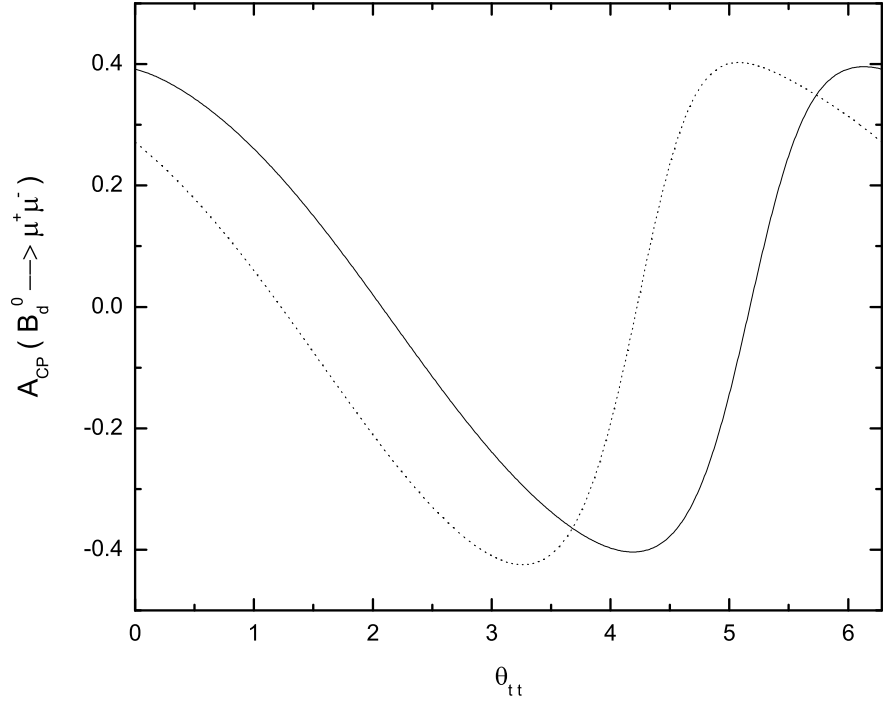,width=13cm}\\
(a)\\
\epsfig{file=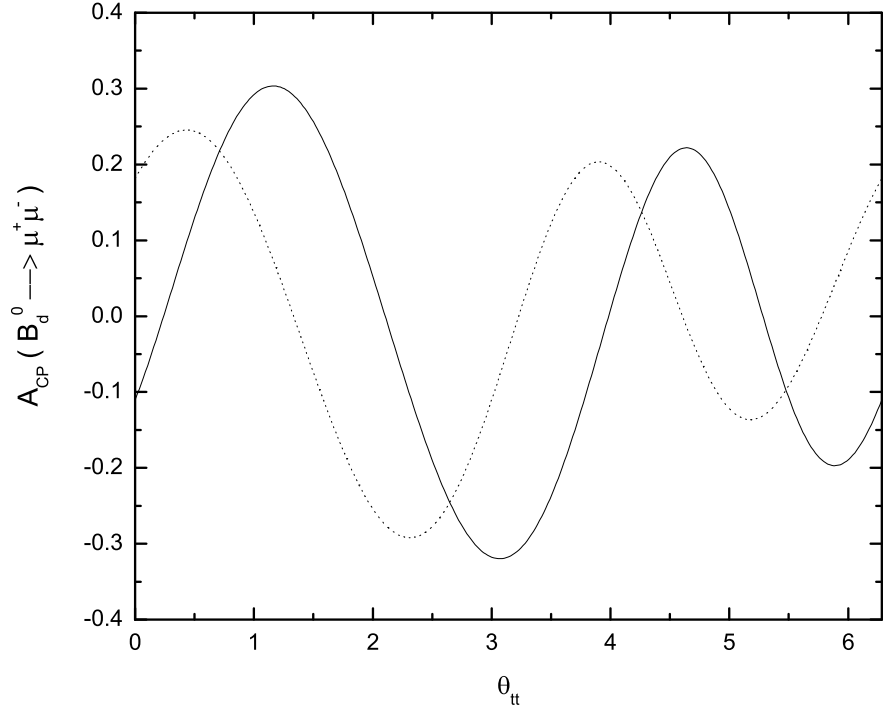,width=13cm}\\
(b)\\
\end{tabular}
\caption{(a) $A_{CP}$ for $B^0_d\rightarrow \mu^+\mu^-$ versus the
CP violating phase $\theta_{tt}$, for $|\lambda_{\mu\mu}|=5$. The
solid line stands for $\theta_{tt}+\theta_{bb}=\pi/2$ , the dotted
line for $\theta_{tt}+\theta_{bb}=\pi/3$. (b)The same as (a)
except for $|\lambda_{\mu\mu}|=50$. } \label{fig:acpmu}
\end{center}
\end{figure}

\end{document}